# Monochromatic CT Image Reconstruction from Current-Integrating Raw Data via Deep Learning

*Wenxiang Cong, Ge Wang*
*Biomedical Imaging Center, Department of Biomedical Engineering,*
*Rensselaer Polytechnic Institute, Troy, NY 12180*

*Abstract* — In clinical CT, the x-ray source emits polychromatic x-rays, which are detected in the current-integrating mode. This physical process is accurately described by an energy-dependent non-linear integral model on the basis of the Beer-Lambert law. However, the non-linear model is too complicated to be directly solved for the image reconstruction, and is often approximated to a linear integral model in the form of the Radon transform, basically ignoring energy-dependent information. This model approximation would generate inaccurate quantification of attenuation image and significant beam-hardening artifacts. In this paper, we develop a deep-learning-based CT image reconstruction method to address the mismatch of computing model to physical model. Our method learns a nonlinear transformation from big data to correct measured projection data to accurately match the linear integral model, realize monochromatic imaging and overcome beam hardening effectively. The deep-learning network is trained and tested using clinical dual-energy dataset to demonstrate the feasibility of the proposed methodology. Results show that the proposed method can achieve a high accuracy of the projection correction with a relative error of less than 0.2%.

## I. INTRODUCTION

Computed tomography (CT) is widely used for medical imaging, allowing visualization and quantification of anatomical and pathological structures with high spatial and contrast resolution. In clinical CT, the energy spectrum of an x-ray source is polychromatic, and the detection of x-rays is operated in the current-integrating mode [1]. This physical process is accurately described by an energy-dependent non-linear integral model on the basis of the Beer-Lambert law. The non-linear model is too complicated to be directly solved for the image reconstruction, and is often approximated to a linear integral model in the form of the Radon transform, basically discarding x-ray energy-dependent information. Because lower energy photons are more easily attenuated than higher energy photons, an x-ray beam would become increasingly harder as it propagates through the patient [2]. Thus, x-ray beams reaching a specific point inside the object from different paths are likely to have different spectra, and are attenuated differently at that point, resulting in attenuation coefficient inconsistencies. This mismatch of computing model to physical model would generate inaccurate quantification of attenuation image and significant beam-hardening artifacts, which has been the subject of serious attention in x-ray medical imaging [2, 3]. Herman proposed a linearization method to transform the measured polychromatic intensity projection data into monochromatic intensity projection data via polynomial fitting to realize monochromatic image reconstruction [3]. Based on the non-linear integral equation, De-Man et. al. developed an iterative maximum-likelihood algorithm [2] to reconstruction the monochromatic linear attenuation coefficient at specific energy level. This method decomposed energy-dependent attenuation coefficients of materials into photoelectric absorption and Compton scattering components, which were correlated to the monochromatic linear attenuation coefficient from prior knowledges. Elbakri et al. described a statistical image reconstruction algorithm based on polychromatic model [4]. It was assumed that the object was segmented a number of nonoverlapping materials from approximately reconstructed image, and the attenuation coefficient was the product of the known energy-dependent mass attenuation coefficient and an unknown density distribution to be reconstructed in the statistical iteration. The iterative algorithms involve a highly nonlinear forward model in the maximum likelihood framework, being a complicated problem of nonlinear optimization at a great computational cost.

Over recent years, deep learning has been extremely successful in image processing, image classification, identification and segmentation [5-10]. In particular, the convolutional neural network (CNN) techniques have become popular for image denoising to transform a low-dose CT image to a high-quality CT image [11, 12]. The deep learning for image reconstruction was also proposed to enhance imaging quality [13]. The deep learning method was developed to learn a regularization transformation and parameters from big data for iterative reconstruction, complying with natural structures of medical images [14]. A novel deep residual network was designed, suppressing artifacts for limited-angle CT image reconstruction [15]. The key idea behind deep learning is not difficult to appreciate. As pointed out in the article [16], "*deep-learning methods are representation-learning methods with multiple levels of representation, obtained by composing simple but non-linear modules that each transform the representation at one level into a representation at a higher, slightly more abstract level. With the composition of enough such transformations, very complex functions can be learned*". In this paper, we propose a deep-learning-based reconstruction algorithm to solve the mismatch problem of computing model to physical model. Our method generates a nonlinear transformation in the form of a deep neural network through a big data training process. This deep learning-based transform would then be capable of correcting measured projection data to accurately match the linear integral model at a target energy level, realizing monochromatic imaging and overcoming the beam hardening effectively. In the next section, a detailed deep-learning-based reconstruction algorithm is proposed for monochromatic x-ray imaging. In the third section, key details are described for the generation of a training dataset, the architecture of our neural network, and the design of the training process. In the fourth section, representative results are reported. In the last section, relevant issues are discussed, key contributions are summarized.

## II. METHODOLOGY

In medical x-ray imaging, the x-ray source generally emits a polychromatic spectrum of x-ray photons, and the x-ray linear attenuation through the object depends on the material composition of the object and the photon energy. After an x-ray beam passes through the object, the x-ray intensity $I(l)$ is measured by a current-integrating detector. The physical process can be exactly described by the non-linear integral model [17]:

$$I(l) = \int_{E_{min}}^{E_{max}} S(E) D(E) \exp\left(-\int_l \mu(r,E) dr\right) dE, \quad (1)$$

where $S(E)$ is the energy spectral distribution of the x-ray source, $D(E)$ is the detection efficiency, and $\mu(r,E)$ is the linear attenuation coefficient at an energy $E$ and a spatial position $r$ along a linear path $l$ through the object. During propagation through the object, the x-ray photons population is statistically attenuated according to the nonlinear equation (1). According to the integral mean value theorem, there is an energy $\varepsilon_l$ for each x-ray path such that the following formula holds:

$$\begin{cases} I(l) = I_0(l) \exp\left(-\int_l \mu(r,\varepsilon_l) dr\right) \\ I_0(l) = \int_{E_{min}}^{E_{max}} S(E) D(E) dE \end{cases}, \quad (2)$$

which is equivalent to

$$\int_l \mu(r,\varepsilon_l) dr = \log\left[\frac{I_0(l)}{I(l)}\right], \quad (3)$$

where $I_0(l)$ is the detected photon number along the path $l$ without any object in the field of view. During a CT scan, x-rays traversing different paths through the object are likely to have different spectra, that is, the different paths correspond to different energy levels $\varepsilon_l$ in Eq. (3). The disagreement in energy levels at different paths is the reason for beam hardening artifacts in a directly reconstructed CT image [18, 19].

The purpose of this paper is to establish a function mapping from deep learning to transform measured projection data $\{\log[I_0(l)/I(l)]\}$ to an monochromatic projection data $\{\log[I_0(l,\varepsilon)/I(l,\varepsilon)]\}$ on a specific energy level $\varepsilon$, which is defined in the detectable energy range. It is not efficient to obtain a direct mapping relation from $\log[I_0(l)/I(l)]$ to $\log[I_0(l,\varepsilon)/I(l,\varepsilon)]$ due to lacking relevant information support between them. While the image $\mu^*(r)$ reconstructed from raw data $\{\log[I_0(l)/I(l)]\}$ contains abundant information in object structure and x-ray attenuation along the associated x-ray paths, a practical method is to map pixel values along an x-ray path on the image $\mu^*(r)$ to the monochromatic projection data $\{\log[I_0(l,\varepsilon)/I(l,\varepsilon)]\}$ at the specific energy level $\varepsilon$ to establish a function relation based on deep learning. Mathematically, the optimization model in the deep learning framework is as follows:

$$\mathcal{M} = \arg\min \sum_{l \in \text{all path set}} \left\| \mathcal{M}(\mu^*(r)|r \in l) - \log\left[\frac{I_0(l,\varepsilon)}{I(l,\varepsilon)}\right] \right\|. \quad (4)$$

If $\mathcal{M}$ is simply taken as a identity transformation, it would generate the uncorrected measurement $\{\log[I_0(l)/I(l)]\}$. If $\mathcal{M}$ is a nonlinear transformation using a sufficiently well-trained neural network, the raw projection data would be effectively corrected to accurately approximate the ideal monochromatic projection data $\log[I_0(l,\varepsilon)/I(l,\varepsilon)]$ at the energy level $\varepsilon$. Once the deep learning-corrected projection dataset is obtained, a monochromatic image $\mu(r,\varepsilon)$ can be reconstructed using a suitable algorithm such as filtered back projection (FBP) or iterative methods based on the line integral model:

$$\int_l \mu(r,\varepsilon) dr = \log\left[\frac{I_0(l,\varepsilon)}{I(l,\varepsilon)}\right]. \quad (5)$$

The reconstruction algorithm based on Eqs. (4) and (5) can realize the monochromatic CT image reconstruction and overcome beam-hardening artifacts efficiently.

## III. NEURAL NETWORK AND TRAINING

A multi-layer perceptron (MLP) neural network is designed to perform deep learning based on Eq. (4). It consists of several layers, including an input layer, hidden layers, and an output layer [20]. Each layer contains multiple nodes, and each node is fully connected to the nodes in a subsequent layer. The pixel values along a path $l$ through a reconstructed image $\mu^*(r)$ are input to nodes of the input layer. The hidden nodes make weighted linear combinations of the outputs from the previous layer. The value at every node is a composition function as follows,

$$y_k = \sum_{s=1}^{N} w_{s,t}^{(n)} \sigma\left[\cdots \sigma\left(\sum_{k=1}^{N}\left(w_{l,k}^{(2)} \sigma\left(\sum_{j=1}^{N} w_{i,j}^{(1)} x_i + b_i^{(1)}\right) + b_k^{(2)}\right)\right)\right], \quad (6)$$

where $\sigma(\cdot)$ is the activation function, and $w_{i,j}^{(n)}$ is the weight for the link between the $i^{th}$ node of the current layer and the $j^{th}$ node of the previous layer, and $b_i^{(n)}$ is the bias at the $i^{th}$ node of the $n^{th}$ layer. For image reconstruction, the output layer only contains a single node yielding a weighted linear combination of all the nodes on the last layer, as shown in Fig.1.

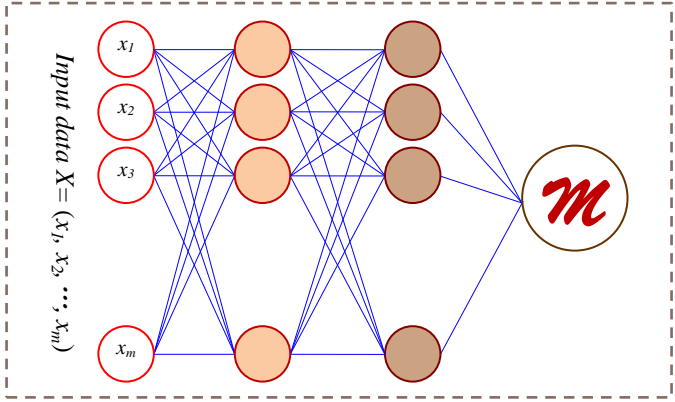

Fig.1. Multi-layer perceptron (MLP) neural network.

The optimal number of hidden layers in the MLP depends on the complexity of the function to be learned. Too few hidden layers would suffer from training errors due to under-fitting. Too many hidden layers would result in over-fitting, reducing convergence speed and network performance. The optimal number of hidden layers can be found iteratively using a heuristic method. The MLP is a supervised learning algorithm that learns a function $\mathcal{M}: R^m \rightarrow R$ by training on a dataset $(X_1,Y_1),(X_2,Y_2),\cdots,(X_N,Y_N)$, where $X$ is conventional CT images, and $Y$ is desired projection data at the specific energy level, $m$ is dimension of the input vector. The variables in the optimization model are nonnegative weight parameters in the neural network. Based on this deep learning procedure, we have the following novel monochromatic image reconstruction Algorithm 1.

**Algorithm 1**

A. *Input: Raw projection data acquired by current-integrating detection mode;*

B. *Perform image reconstruction from raw data using FBP;*

C. *Learn a non-linear map from big dataset for the correction of projection data:*

$$\mathcal{M} = \arg\min \sum_{l \in all\ path\ set} \left\| \mathcal{M}\left(\mu^*(r) \middle| r \in l\right) - \log\left[\frac{I_0(l,\varepsilon)}{I(l,\varepsilon)}\right] \right\|$$

D. *Reconstruct an image from corrected projections;*

E. *Output: Reconstructed monochromatic image.*

The neural network for tomographic imaging must be trained with a large dataset to optimize its performance. The training dataset are conventional CT images and their corresponding monochromatic projection data. To ensure accuracy, uniqueness and stability of the solution, the training dataset should have a sufficiently high quantity of images with well-diversified features. The monochromatic projection data can be obtained from dual-energy CT. Existing dual-energy CT scanners include the source kVp-switching (GE), double-layer detection (Philips), dual-source gantry (Siemens), and two-pass scanning (Toshiba). Dual-energy CT is able to reconstruct monochromatic images of an object from two projection datasets generated with two distinct x-ray energy spectra, and provides more accurate attenuation quantification than conventional CT with a single x-ray energy spectrum. In the diagnostic energy range, x-ray energy-dependent attenuation can be approximated as a combination of photoelectric absorption and Compton scattering. Dual-energy CT provides complete energy-dependent information under the assumption of two basis materials. Recently, a new projection decomposition method has been proposed for the image reconstruction in dual-energy CT. This method combines both an analytical algorithm and a single-variable optimization method to solve the non-linear polychromatic x-ray integral model, and can achieve a low computational cost, accurate quantification of photoelectric absorption and Compton scattering components, realizing monochromatic image reconstruction and material decomposition [21].

IV. EXPERIMENT RESULTS

We performed numerical tests to evaluate the proposed deep-learning-based algorithm for monochromatic CT imaging. A set of clinical dual-energy CT dataset of the human abdomen from Ruijin hospital in Shanghai, China, collected on a GE Discovery CT750 scanner, were used to perform the training of neural network. The dataset included monochromatic images at 8 narrow energy bins. Each energy bins includes 181 slices with 512×512 pixels. Ideal monochromatic image was taken in the fourth energy channel. An x-ray spectrum was generated using an open source software package from Siemens website (https://health.siemens.com/booneweb/index.html) for the CT image reconstruction from the monochromatic image dataset.

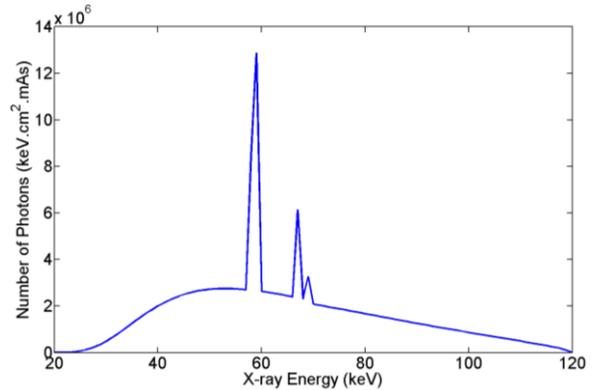

Fig. 2. Energy spectral distribution of the x-ray source simulated using the public software (https://health.siemens.com/booneweb/index.html). The energy spectrum generated from the x-ray tube (120 kvp) was filtered by Aluminum of 1mm thickness and Copper of 0.3mm thickness.

In the setting of the x-ray imaging, the field of view (FOV) is of a 25 cm radius, and the radius of the scanning trajectory is 53.85cm. Source-to-detector distance is 94.67cm. 540 projections are uniformly acquired over a 360-degree angular range. 765 detector elements with 0.1024cm pitch were equiangular distributed on a projection view. The neural network was designed to have four layers, one input layer, two hidden layers, and one output layer. The sigmoid activation function was used in the network. The neural network was trained using the Adam optimization algorithm. The training procedure was programmed in Python in the TensorFlow framework, running 30 iterations on a computer with a NVIDIA Titan XP GPU of 12 GB memory. The optimization of the neural network has an excellent convergent performance, and its cost function is basically monotonously decreasing in the big data learning. Well-trained neural network is able to achieve a high accuracy of projection correction with a relative error less than 0.2%, here the definition of relative error is

$$\left| \frac{True\ projection - Corrected\ projection}{True\ projection} \right|.$$

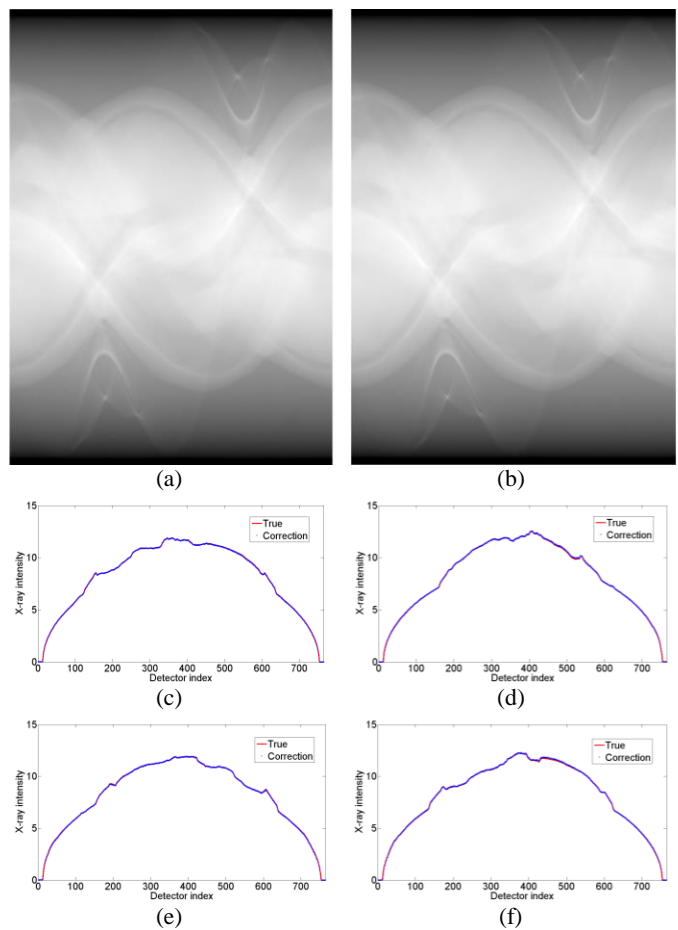

Fig.3. Sinogram corrections. (a) The ground truth sinogram, and (b) the reconstructed sinogram. (c-f) The comparisons of the first, $100^{th}$, $300^{th}$ and $500^{th}$ projection views, respectively.

To assess the performance of the trained network based on Algorithm 1 listed in Table 1, conventional CT images were input to the well trained network to produce corrected projection data. Fig. 3 presents the comparison of reconstructed sinogram to ground truth sinogram, and the comparison for several view projections. Then, the image reconstruction was performed using FBP from the corrected projection data. Fig. 4 shows a comparison between the reconstructed image and the ground truth for the representative example of monochromatic image reconstruction.

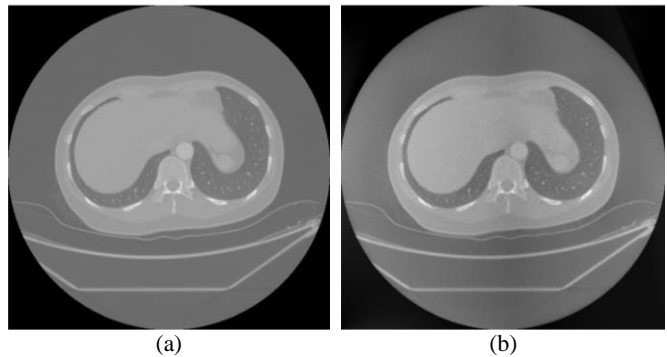

Fig. 4. Monochromatic image reconstruction. (a) The ground truth image, and (b) the image reconstructed from deep learning based reconstruction method.

## V. DISCUSSIONS AND CONCLUSION

The energy-dependent non-linear integral equation on the basis of the Beer-Lambert law is an accurate physical model for x-ray CT, and is too complicated to be directly solved for the image reconstruction. The linear integral equation in the form of the Radon transform is an ideal computational model, and has an analytical inverse formula, such as filtered backprojection (FBP), producing accurate and stable solution, while it is only an approximation to the non-linear integral equation, basically ignoring energy-dependent information, and generating inaccurate quantification of attenuation image and significant beam-hardening artifacts. The mismatch of computing model to physical model is also the most fundamental problem in the x-ray computed tomographic imaging. The proposed deep-learning-based sinogram correction method successfully addresses the issue. The optimization of the neural network has an excellent convergent performance in the big data learning process, and achieves a high accuracy of the projection correction with a relative error of less than 0.2%. This method learns a nonlinear transformation from big data to correct measured projection data to accurately match the linear integral model, realizing monochromatic imaging and overcome beam hardening effectively. The proposed method is applicable to biomedical imaging, nondestructive testing, security screening, and other applications.

**Acknowledgments:** This work was supported by National Institutes of Health Grant NIH/NIBIB R01 EB016977 and U01 EB017140.